\documentclass[sigconf, nonacm]{acmart}

\usepackage{algorithm}
\usepackage{algpseudocode}
\usepackage{xspace}
\usepackage{xcolor, listings}

\newcommand{\ie}{\emph{i.e.,}\xspace}
\newcommand{\eg}{\emph{e.g.,}\xspace}

\newcommand\vldbdoi{10.14778/3611540.3611608}
\newcommand\vldbpages{4010-4013}
\newcommand\vldbvolume{16}
\newcommand\vldbissue{12}
\newcommand\vldbyear{2023}

\newcommand\vldbtitle{\shorttitle} 

\newcommand\vldbpagestyle{empty}

\begin{document}
\title{\textsc{VisualNeo}: Bridging the Gap between Visual Query Interfaces and Graph Query Engines}

\author{Kai Huang}
\affiliation{%
  \institution{Hong Kong Univ. of Sci. \& Tech.}
}
\email{ustkhuang@ust.hk}

\author{Houdong Liang$^{\S}$}
\affiliation{%
  \institution{Hong Kong Univ. of Sci. \& Tech.}
}
\email{hliangam@connect.ust.hk}

\author{Chongchong Yao$^{\S}$}
\affiliation{%
  \institution{Hong Kong Univ. of Sci. \& Tech.}
}
\email{cyaoad@connect.ust.hk}

\author{Xi Zhao}
\affiliation{%
  \institution{Hong Kong Univ. of Sci. \& Tech.}
}
\email{xzhaoca@connect.ust.hk}

\author{Yue Cui}
\affiliation{%
  \institution{Hong Kong Univ. of Sci. \& Tech.}
}
\email{ycuias@cse.ust.hk}

\author{Yao Tian}
\affiliation{%
  \institution{Hong Kong Univ. of Sci. \& Tech.}
}
\email{ytianbc@cse.ust.hk}

\author{Ruiyuan Zhang}
\affiliation{%
  \institution{Hong Kong Univ. of Sci. \& Tech.}
}
\email{zry@ust.hk}

\author{Xiaofang Zhou}
\affiliation{%
  \institution{Hong Kong Univ. of Sci. \& Tech.}
}
\email{zxf@ust.hk}

\begin{abstract}
Visual Graph Query Interfaces (VQIs) empower non-programmers to query graph data by constructing visual queries intuitively. Devising efficient technologies in Graph Query Engines (GQEs) for interactive search and exploration has also been studied for years. However, these two vibrant scientific fields are traditionally independent of each other, causing a vast barrier for users who wish to explore the full-stack operations of graph querying. In this demonstration, we propose a novel VQI system built upon Neo4j called \textsc{VisualNeo} that facilities an efficient subgraph query in large graph databases. \textsc{VisualNeo} inherits several advanced features from recent advanced VQIs, which include the data-driven \textsc{gui} design and canned pattern generation. Additionally, it embodies a database manager module in order that users can connect to generic Neo4j databases. It performs query processing through the Neo4j driver and provides an aesthetic query result exploration.
\end{abstract}

\maketitle

\pagestyle{\vldbpagestyle}
\begingroup\small\noindent\raggedright\textbf{PVLDB Reference Format:}\\
Kai Huang, Houdong Liang, Chongchong Yao, Xi Zhao, Yue Cui, Yao Tian, Ruiyuan Zhang, Xiaofang Zhou. \vldbtitle. PVLDB, \vldbvolume(\vldbissue): \vldbpages, \vldbyear.\\
\href{https://doi.org/\vldbdoi}{doi:\vldbdoi}
\endgroup
\begingroup
\renewcommand\thefootnote{}\footnote{\noindent
This work is licensed under the Creative Commons BY-NC-ND 4.0 International License. Visit \url{https://creativecommons.org/licenses/by-nc-nd/4.0/} to view a copy of this license. For any use beyond those covered by this license, obtain permission by emailing \href{mailto:info@vldb.org}{info@vldb.org}. Copyright is held by the owner/author(s). Publication rights licensed to the VLDB Endowment. \\
\raggedright Proceedings of the VLDB Endowment, Vol. \vldbvolume, No. \vldbissue\ %
ISSN 2150-8097. \\
\href{https://doi.org/\vldbdoi}{doi:\vldbdoi} \\
}\addtocounter{footnote}{-1}\endgroup



\section{Introduction}

Visual graph query interfaces (VQI) enable non-experts to compose graph queries effortlessly without writing any textual query language, which broadens the usability of graph querying frameworks. Consequently, numerous academic and commercial frameworks for querying large graph databases adopt VQIs for composing subgraph queries. For example, \textit{PubChem} \cite{pubchem} provides a VQI for researchers in the chemistry domain to perform chemical compound searches. The \textsc{gui} of \textit{PubChem} \cite{pubchem} includes a label panel containing a set of chemical elements and a pattern panel with carboxyl groups, a benzene ring, etc. However, the contents of the panel are typically chosen manually by domain experts. 


To address this challenge, data-driven VQI design has attracted considerable attention in recent years \cite{yuan2022playpen, huang2022vincent, bhowmick2020aurora, huang2017picasso}. Given a graph database $D$, data-driven VQIs automatically populate panels (\eg label panel, pattern panel) of the \textsc{gui} from $D$. While the label and property panels can be populated without many difficulties by traversing the underlying databases, selecting useful patterns is an NP-hard problem \cite{huang2019catapult, yuan2021towards,huang2021midas}. A few algorithms have been proposed to conduct the selection of these patterns \cite{yuan2021towards, huang2023ted}. TATTOO framework \cite{yuan2021towards} performs data-driven selection of canned patterns for large networks while TED framework \cite{huang2023ted} works for a large collection of small- or medium-size data graphs. Several data-driven VQI systems have been developed as well \cite{yuan2022playpen, huang2022vincent}.

Despite the amount of progress made in the development of data-driven VQIs, a potential direction is overlooked, which is the connection between VQI and Graph Query Engines (GQEs). A GQE acts as an abstraction layer between the user application and the database. It receives database operations in the form of query language and outputs specific data to the user application. Therefore, VQI complements GQE in the way that it provides an easy-to-use interface for non-professional users to construct their queries, which are subsequently sent to GQE for processing and result exploration. Traditional data-driven VQI design focus on efficient subgraph query processing \cite{jin2012prague} and data-driven selection of canned patterns \cite{yuan2022playpen, huang2022vincent}, but ignores the potential of GQEs in terms of abilities to process queries.

In this demonstration, we present a novel data-driven visual subgraph query system called \textsc{VisualNeo}. The system is built upon Neo4j \cite{neo4j}, a popular graph query engine. It possesses a \textit{Database Manager} module with which users can connect to local or remote Neo4j graph databases by providing authentication information. It supports generic Cypher query processing via Neo4j driver, which includes nodes, relationships, labels, and properties. Therefore, \textsc{VisualNeo} builds a bridge between VQI and GQE, introducing a new direction for these two scientific fields.

\textsc{VisualNeo} also inherits several state-of-the-art features from recent data-driven VQI designs. First, it generates a set of diversified patterns called \textsc{TED} patterns (\ie \underline{T}op-k \underline{E}dge-\underline{D}iversified patterns) \cite{huang2023ted}, which summarize the characteristics of the underlying database and thus facilitate the query formulation. In particular, TED patterns have a theoretical guarantee of the edge coverage approximation ratio and its generation process requires limited memory. Second, it embodies an aesthetic query results explorer which adopts the Fruchterman-Reingold algorithm \cite{fruchterman1991graph} to display the retrieved results.

\textsc{VisualNeo} embraces innovative features as well. To ensure user-friendliness, VisualNeo provides adequate support during the query formulation process. It displays metadata information of the underlying database and provides real-time query translation to guide users to perform exploratory searches where users are unsure about their initial goals or ways to achieve their goals.

\begin{figure}
	\centering
	\includegraphics[width=1\linewidth,height=4.5cm]{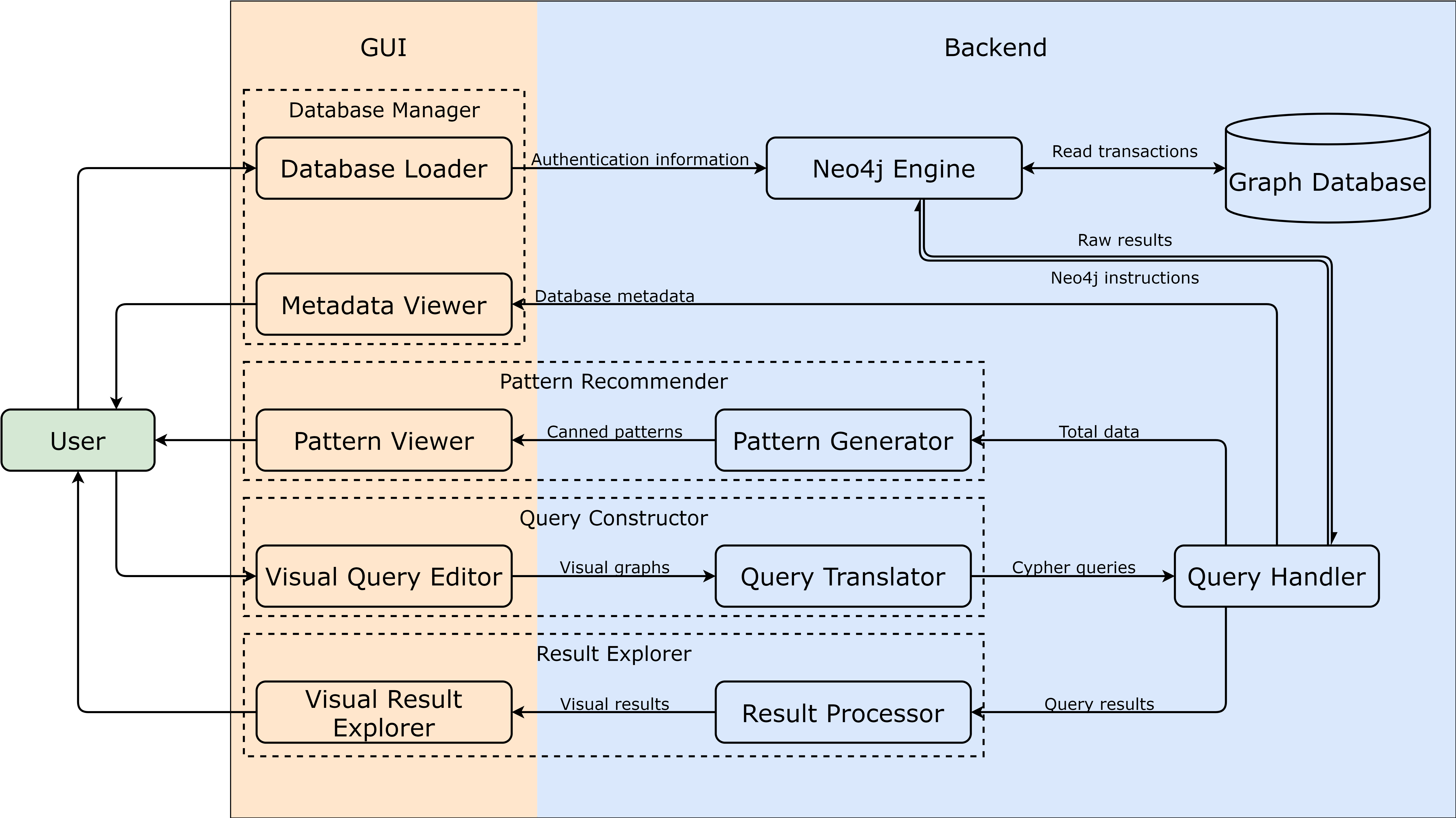}
	\caption{The architecture of \textsc{VisualNeo}.}\label{fig:architecture}
\end{figure}

\section{System Architecture}
Figure \ref{fig:architecture} shows the architecture of \textsc{VisualNeo}. It consists of five modules, \textit{Database Manager}, \textit{Pattern Recommender}, \textit{Query Constructor}, \textit{Query Handler}, and \textit{Result Explorer}. The \textit{Database Manager} module first establishes a connection to a graph database server with user authentications and displays its metadata obtained by the \textit{Query Handler} module. The \textit{Query Handler} module also exports the whole database such that the \textit{Pattern Recommender} module can then utilize METIS \cite{karypis1997metis} (for graph partitioning) and TED \cite{huang2023ted} (for pattern generation) to produce diversified and high-coverage patterns. With the aid of these two modules, the \textit{Query Constructor} module enables the user to form visual query graphs effortlessly. To integrate with the GQE (\ie Neo4j), the query graphs are further translated into formal query languages and fed to the \textit{Query Handler} module. Next, the \textit{Query Handler} module instructs the GQE to execute read transactions and retrieve the desired results. Finally, the \textit{Result Explorer} module converts the unprocessed results into navigable visual graphs for the user to investigate. 

\begin{figure}
	\centering
	\includegraphics[width=1\linewidth,height=4.2cm]{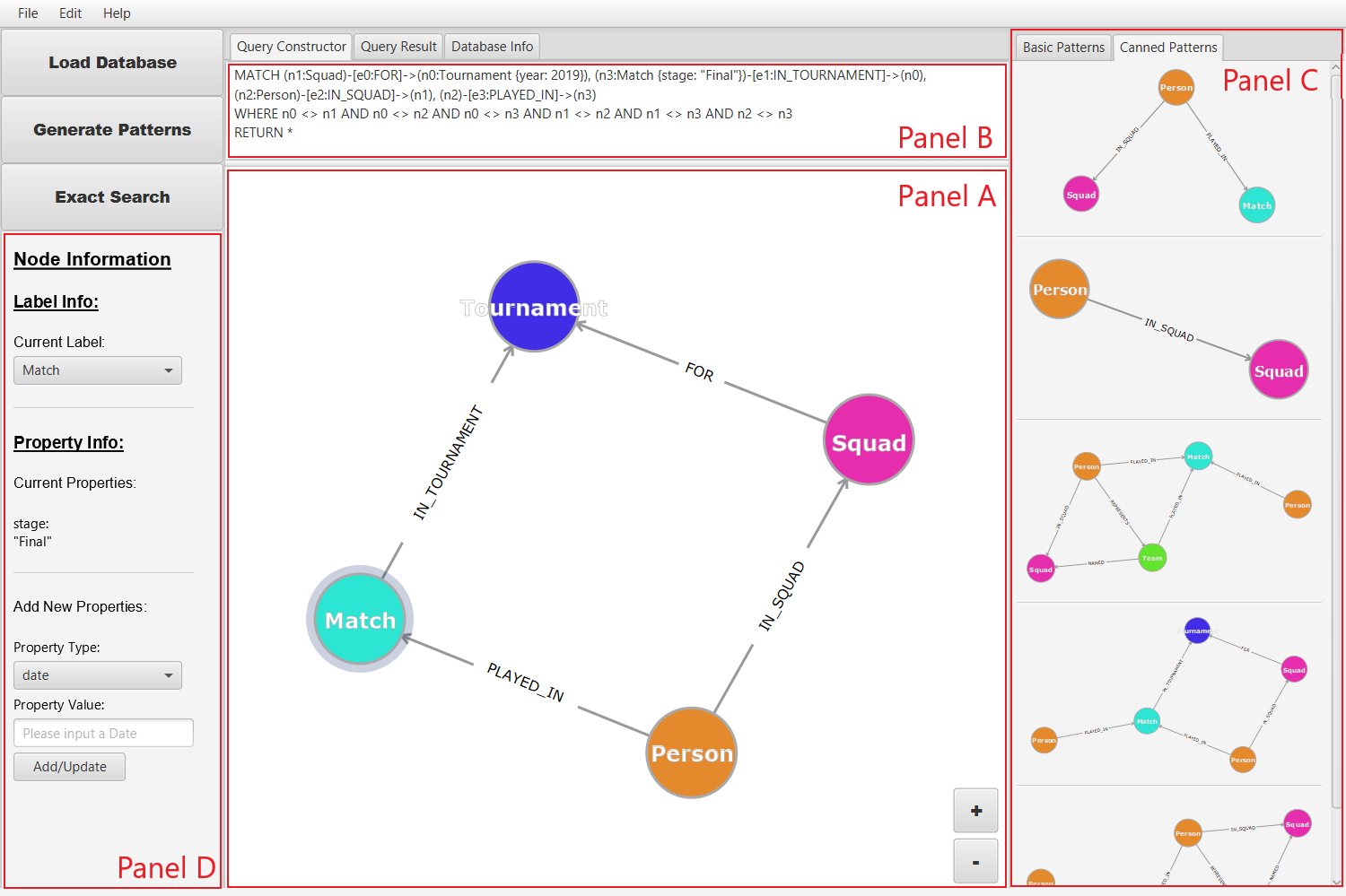}
	\caption{The \textit{Query Constructor panel}.}\label{fig:construct}
\end{figure}

\begin{figure}
	\centering
	\includegraphics[width=1\linewidth,height=4.3cm]{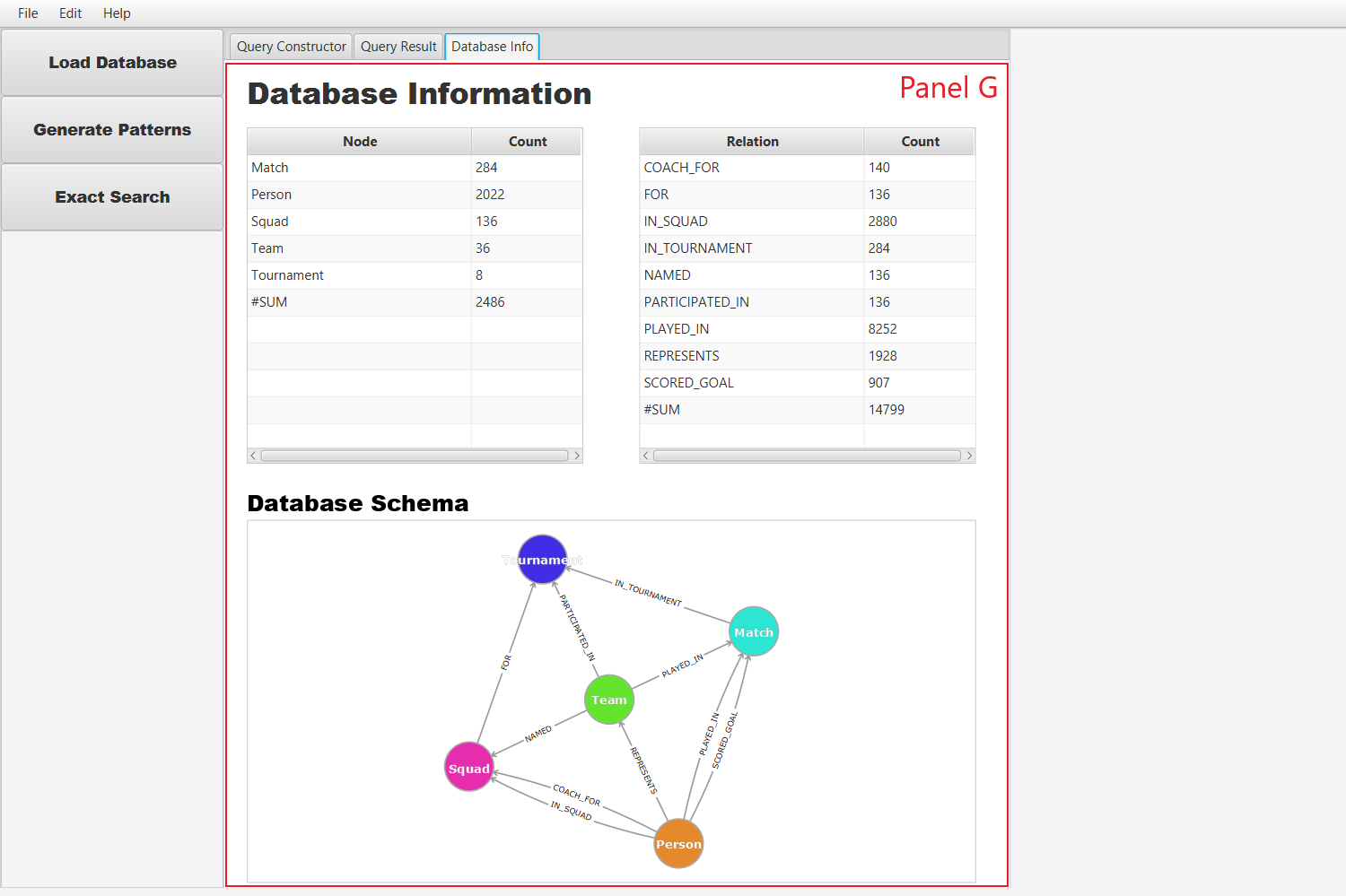}
	\caption{The \textit{Database Info panel}.}\label{fig:info}
\end{figure}

\textbf{\underline{Database Manager module.}}
This module establishes a connection to a graph database server and displays its metadata. When the user clicks the "Load Database" button shown in Figure \ref{fig:construct} and specifies the authentication information, the \textit{Database Manager} module sends a connection request to the server. \textsc{VisualNeo} is compatible with all standard Neo4j servers and only requires read authority. After connection, the \textit{Query Handler} module automatically executes a sequence of queries to retrieve database metadata, including node/relationship counts, node/relationship labels, node/relationship property keys \& data types, and the schema graph that defines the topology among different classes of nodes and relationships. These queries either directly access the Neo4j count store or invoke database procedures. Therefore, such searches have $O(1)$ time complexity for each class/property. The metadata is then displayed in the Database Information panel (\textit{Panel G}) shown in Figure \ref{fig:info} and is also used to facilitate query formulation by providing label and property constraints in the \textit{Query Constructor} module.

\textbf{\underline{Pattern Recommender module.}}
This module generates TED patterns and populates the canned pattern panel when the user clicks the “Generate Pattern” button shown in Figure \ref{fig:construct} and specifies the constraints of desired patterns. 

\begin{definition}[\textbf{Top-k Edge-Diversified Patterns (TED patterns)}]
	{\em
		Given a graph $G$ and an integer $k$, the top-$k$ edge-diversified patterns is the set of  $k$ connected subgraphs $\mathcal{P} = \{p_1, p_2,…,p_i,…p_k\}$  in $G$ such that the total coverage of $\mathcal{P}$ over $G$ (denoted by $|Cov(\mathcal{P}, G)|$), \ie
		$|\cup_i  Cov(p_i, G)|$, 
		is maximized,  where  $Cov(p_i, G) $ is the cover set  of $p_i$ over $G$ (\ie the covered edges of $p_i$ over $G$). 
		\/}
\end{definition}

The Pattern Recommender module consists of two components: METIS graph partitioning \cite{karypis1997metis} and TED pattern generation \cite{huang2023ted}. METIS partitions the large network into a large collection of small- or medium-sized graphs (\eg tens of nodes per graph). Next, the TED algorithm greedily searches for a set of patterns with maximum edge coverage.   Given a set of partitions $D = \{G_1,G_2,...G_n\}$ and an integer $k$, the TED algorithm first enumerates all 1-sized subgraphs (\ie edges) and appends them into the set $\mathcal{P}$. Then, an iterative process is performed to enumerate $\tau$-sized ($\tau \geq 2$) subgraphs by performing the right-most extension \cite{yan2002gspan}  and appends them into the set $\mathcal{P}$. When the size of $\mathcal{P}$ is $k$ and a new subgraph $g$ is generated,  a swapping-based process is developed to determine if $g$ should be swapped into $\mathcal{P}$.  In particular, it first calculates the loss score (\ie decreased coverage if $p$ is swapped out) for $p\in \mathcal{P}$ and then records the pattern $p_t$ and its pattern score $\textsc{Score}_L$ such that  $p_t$  has the minimum loss score. The benefit score  (\ie increased coverage if $g$ is swapped in) $\textsc{Score}_B$ of $g$ is also recorded.  The subgraph $g$ is considered as a promising candidate and swapped into $\mathcal{P}$  if 
	$\textsc{Score}_B > (1+\alpha)\textsc{Score}_L  + (1-\alpha)|Cov(\mathcal{P}, D)|/k$ is satisfied,
where $\alpha \in [0,1]$ is a \textit{swapping threshold} for balancing loss score $\textsc{Score}_L$ and the average coverage of patterns in $\mathcal{P}$. The approximation ratio (w.r.t., total coverage) of patterns  $\mathcal{P}$  is bounded by $|Cov(\mathcal{P},D)|$$/$$|Cov(\mathcal{P}_{opt},D)| \geq \frac{1}{4}$ where $\mathcal{P}_{opt}$ is the optimal solution, which can be obtained by enumerating all subgraphs and generating all possible combinations of $k$ subgraphs.

\textbf{\underline{Query Constructor module.}}
This module provides the user with a editor for visual query formulation and subsequently translates the visual graphs into Cypher queries. As shown in Figure \ref{fig:construct}, it comprises five components: the \textit{Graph Editor} panel (\textit{Panel A}), the \textit{Query View} panel (\textit{Panel B}), the \textit{Pattern View} panel (\textit{Panel C}), and the \textit{Element Constraint} panel (\textit{Panel D}).



\begin{itemize}
\item Elements can be selected/added/removed in \textit{Panel A} using mouse-keyboard operations, with navigation and zoom adjustment possible through dragging and scrolling.
\item \textit{Panel B} provides a real-time translation of the visual graphs in \textit{Panel A} to Cypher statements.
\item \textit{Panel C} displays TED patterns generated by the \textit{Pattern Recommender} module and basic patterns applicable to universal databases, which can be added to \textit{Panel A} through drag-over operations.
\item Elements' labels and properties can be viewed and edited in \textit{Panel D}.
\item Upon completing the graph, clicking the "Exact Search" button in Figure \ref{fig:construct} sends the translated Cypher query to the Query Handler module.
\end{itemize}

\begin{figure}
	\centering
	\includegraphics[width=1\linewidth,height=1.2cm]{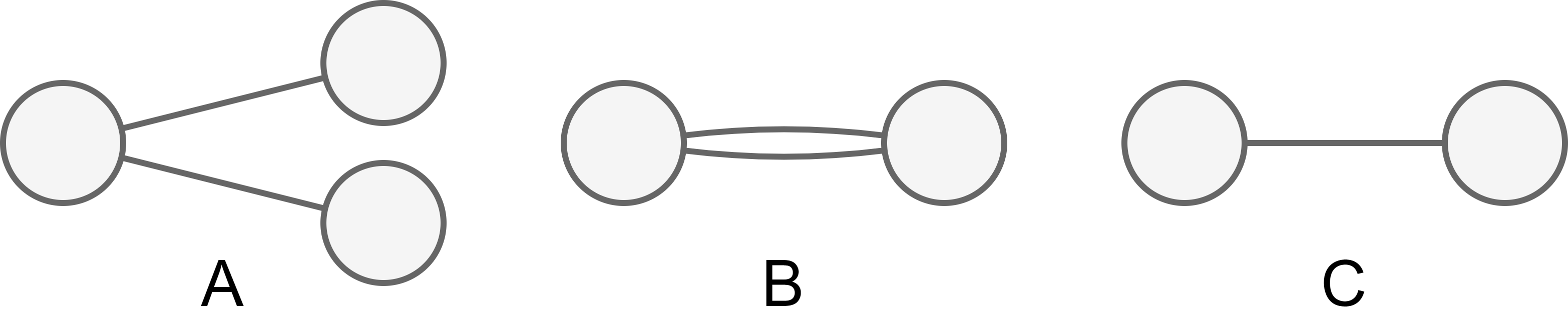}
	\caption{Indistinguishable patterns without node isomorphism (A and B) or relationship isomorphism (A and C).}\label{fig:patterns}
\end{figure}

Following each step of building a visual query, the resulting query is automatically translated into a formal Cypher query by traversing all relevant relationships. For each relationship, the translator generates a corresponding line in the \texttt{MATCH} clause in the form of \texttt{(startNode)-[thisRelation]-(endNode)}. Nevertheless, this approach is susceptible to isomorphic issue, meaning that the same node/relationship may be returned more than once for each matching record. For instance, if a user creates a visual query in the form of \texttt{(n2)-[r1]-(n1)-[r2]-(n3)} (Pattern A in Figure \ref{fig:patterns}), it may result in a mismatch with Patterns B or C if the translated query lacks appropriate isomorphic constraints. This highlights the need for additional measures to ensure accurate and reliable query translation. In practice, Neo4j Cypher utilizes relationship isomorphism for path matching, thereby preventing the same relationship from being returned more than once within a single result record. However, it does not assert node isomorphism. While this matching mechanism may assist programmers in managing complex queries, inexperienced users may be misled, as demonstrated by the fact that Pattern B in Figure \ref{fig:patterns} is considered a valid match for Pattern A. Most existing Cypher-oriented VQI (\eg Popoto.js \cite{popotojs}), do not handle node isomorphism because they do not connect to a GQE as a backend. To address this issue, we add additional inequality constraints on each node pair in the \texttt{WHERE} clause to ensure node isomorphism. To further enhance query efficiency, we eliminate trivial inequalities before feeding them into the constraints. For instance, if two nodes have distinct labels or properties, the corresponding inequality can be removed.
\color{black}

\textbf{\underline{Query Handler module.}}
This module’s features include executing Cypher queries via the Neo4j driver, receiving query results, extracting desired data from returned records, converting raw data into Java-typed data, and boxing data into proper containers for further processing. All sessions and transactions are read-only to avoid modification and undesired authentication errors. Besides, consecutive transactions (e.g., metadata queries) are bundled into a single session to enhance efficiency.

\begin{figure}
	\centering
	\includegraphics[width=1\linewidth,height=4.4cm]{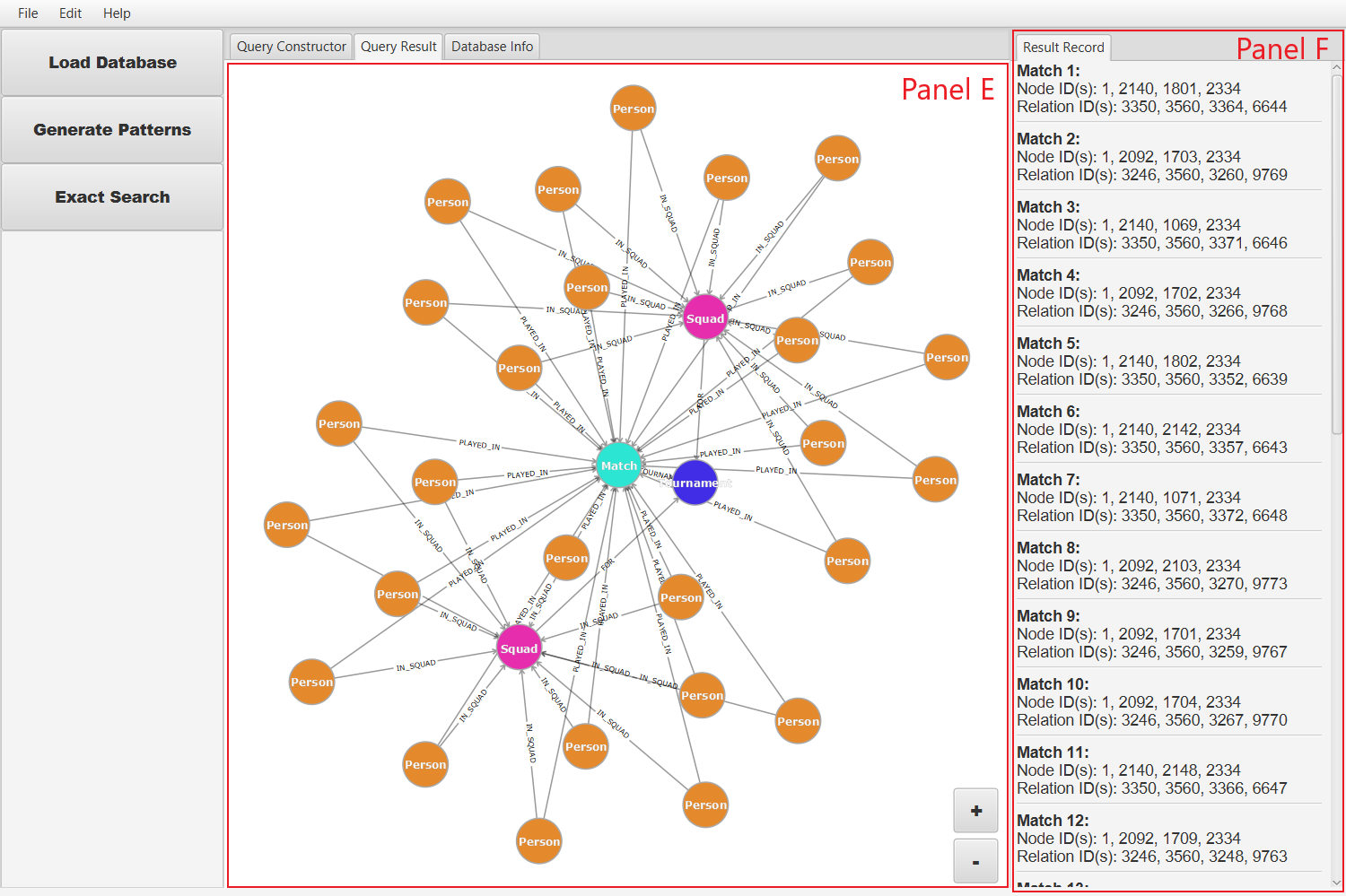}
	\caption{The \textit{Query Result panel}.}\label{fig:result}
\end{figure}

\textbf{\underline{Result Explorer Module}}
This module is responsible for processing query results and presenting them in an aesthetically pleasing and navigable manner. It has two components: \textit{Panel E} displays the result graph, while \textit{Panel F} lists all matching records. With a click on an item in \textit{Panel F}, \textit{Panel E} immediately navigates to the corresponding pattern and highlights it.

In cases where the query involves nodes with high centrality or relationships with high betweenness, it is likely that the result contain duplicate elements. This occurrence of redundant information can cause high data transfer traffic and memory cost at the local device. To avoid such inefficiencies, we generate an ID reference list with Cypher for each query and only keep the information of distinct elements. This approach ensures that the final result is devoid of any information loss, and simultaneously prevent unnecessary data transfer or memory usage at the local device.

To arrange the results in an orderly manner, we employ a modified version of the Fruchterman-Reingold (FR) force-directed graph layout algorithm \cite{fruchterman1991graph}. First, we prune self-loops and retain only one relationship between each unordered pair of nodes to avoid unnecessary computations and excessive proximity between heavily connected nodes. Furthermore, we specify a constant optimal distance between connected node pairs and remove boundaries around the graph to enhance visual effects and reduce computation cost. Additionally, we establish a maximum distance on repulsive forces to prevent disjoint subgraphs from repelling each other to infinity. Finally, we center the centroid of all nodes at the origin after the force simulation to cancel the universal offset. The additional $O(|V|)$ computation cost for computing the centroid once for each graph is negligible compared to the overall $O(K(|V|^2+|E|))$ time complexity, where $G=(V, E)$ is the graph and $K$ is the number of simulation iterations.
\color{black}

\section{Related Systems and Novelty}
Graph Query Engines (GQEs) such as Neo4j \cite{neo4j} assume that a user has programming and debugging
expertise to formulate queries correctly with query languages (\eg Cypher). This
assumption makes it harder for non-programmers to take advantage of a graph querying framework.
Although existing Visual Query Interfaces (VQIs) such as \textsc{PLAYPEN} \cite{yuan2022playpen} and \textsc{VINCENT} \cite{huang2022vincent} can alleviate this problem by enabling users to visually formulate queries,   they ignore the potential of GQEs in terms of abilities to process queries. In contrast, \textsc{VisualNeo} has the following advantages. First, \textsc{VisualNeo}  supports not only visual query formulation but also efficient graph query processing by leveraging the strength of GQEs. That is, it bridges the gap between VQIs and GQEs. Second, the patterns generated for visual query formulation can achieve a guaranteed approximation ratio of edge coverage and the generation process requires limited memory. Third, in contrast to \textsc{VINCENT}’s hierarchical layout, \textsc{VisualNeo} utilizes the Fruchterman-Reingold algorithm \cite{fruchterman1991graph} to display a force-directed graph drawing, which better suits general databases whose structure is more uncertain. Lastly, \textsc{VisualNeo} supports queries in an attributed graph and query translation, while both \textsc{PLAYPEN} and \textsc{VINCENT} only support queries in a simple graph.

Existing VQI libraries for Neo4j such as Popotojs \cite{popotojs} and tools such as Graphileon \cite{graphileon} only support  iterative constructions of edges one-at-a-time (\ie edge-at-a-time mode), while \textsc{VisualNeo} enables a user to construct multiple nodes and edges in a subgraph query by performing a single click-and-drag action (\ie pattern-at-a-time mode) and thus facilitate efficient query formulations.

\section{Demonstration Overview}
\textsc{VisualNeo} is implemented in Java JDK 17 and JavaFX 19. In the demonstration, it will be loaded with a few real-world databases (\eg Women's World Cup 2019) from Neo4j Sandbox \cite{sandbox}. Example query graphs that can be constructed using patterns will be presented for formulation. Users can write their own ad hoc queries through our visual query editor as well. The key objective of the demonstration is to lead users through the full-stack operations of graph querying with the aid of data-driven VQIs and GQEs. In particular, it enables the audience to experience the following:

\textbf{Scenario 1: Database loading and metadata information display.} The “Load Database” button shown in Figure \ref{fig:construct} enables the audience to connect to local or remote databases. After the database is loaded, the metadata information can be viewed in Figure \ref{fig:info}. Users can obtain a macroscopic understanding of the underlying database by looking into the displayed node/relationship label table and schema graph.

\textbf{Scenario 2: Data-driven visual query formulation.} Through the “Generate Pattern” button shown in Figure \ref{fig:construct}, users can set the hyperparameters for TED frameworks and launch the generation of the TED patterns. The TED patterns will then be displayed in \textit{Panel C}. In the process of query formulation, users can create nodes/relationships directly or drag-and-drop basic patterns or TED patterns from \textit{Panel C}. 

\textbf{Scenario 3: Real-time query translation.} \textsc{VisualNeo} supports real-time query translation for users’ reference. In the process of visual query formulation, users can get familiar with formal query languages by observing the translated Cypher query (see \textit{Panel B}) of the graph query (see \textit{Panel A}). Consequently, users are able to construct their desired queries effortlessly.  
\color{black}

\textbf{Scenario 4: Aesthetic query result exploration.} After users click the “Exact Search” button shown in Figure \ref{fig:construct}, \textsc{VisualNeo} will start the query processing, display the result graphs in a force-directed way in \textit{Panel E}, and enable users to iterate through the query results using matching records in \textit{Panel F}.

A demonstration video is publicly available at \url{https://youtu.be/th0LqEK-S3s}.


\begin{acks}  
The work was conducted in the JC STEM Lab of Data Science Foundations funded by The Hong Kong Jockey Club Charities Trust. 
\end{acks}

\bibliographystyle{ACM-Reference-Format}

\end{document}